# EUROPEAN HISTORICAL EVIDENCE OF THE SUPERNOVA OF AD 1054

# COINS OF CONSTANTINE IX AND SN 1054

**Miroslav D. Filipović[1*], Jeffrey L. Payne[1], Thomas Jarrett[2], Nick F.H. Tothill[1], Evan J. Crawford[1], Dejan Urošević[3], Giuseppe Longo[4], Jordan D. Collier[5], Patrick J. Kavanagh[4], Christopher Matthew[1] and Miro Ilić[6]**

[1] Western Sydney University, School of Science, Locked Bag 1797, Penrith, NSW 2751, Australia
[2] University of Cape Town, The Inter-University Institute for Data Intensive Astronomy (IDIA), Department of Astronomy, Private Bag X3, Rondebosch 7701, South Africa
[3] University of Belgrade, Faculty of Mathematics, Department of Astronomy, Studentski trg 16, 11000 Belgrade, Republic of Serbia
[4] University Federico II, Department of Physics, Via Cinthia 6, I-80126 Napoli, Italy
[5] Dublin Institute for Advanced Studies, School of Cosmic Physics, 31 Fitzwilliam Place, Dublin 2, Ireland
[6] Trebinje Astronomical Association, Republike Srpske 15, Trebinje, 89101, Republic Srpska, Bosnia and Herzegovina

## Abstract

We investigate a possible depiction of the famous SN 1054 event in specially minted coins produced in the Eastern Roman Empire in 1054 A.D. On these coins, we investigate if the head of the Emperor, Constantine IX, might represent the Sun with a bright 'star' on either side - Venus in the east and SN 1054 in the west, perhaps also representing the newly split Christian churches. We explore the idea that the eastern star represents the stable and well-known Venus and the Eastern Orthodox Church, while the western star represents the short-lived 'new star' and the 'fading' Western Catholic church. We examined 36 coins of this rare Constantine IX Class IV batch. While no exact date could be associated to any of these coins, they most likely were minted during the last six months of Constantine IX's rule in 1054. We hypothesise that the stance of the church concerning the order of the Universe, as well as the chaos surrounding the Great Schism, played a crucial role in stopping the official reporting of an obvious event in the sky, yet a dangerous omen. A temporal coincidence of all these events could be a reasonable explanation as well.

*E-mail: m.filipovic@westernsydney.edu.au

## 1. Introduction

[1] and [2] surveyed the records of European history and culture from around 1054 A.D. to understand the absence of historical records for SN 1054. We asked if SN 1054 was seen in the skies above Europe and did it have an impact on the people who saw it. We have reviewed and analysed several factors that could account for the lack of European records of SN 1054. Possible explanations for this mystery include political/religious reasons, or perhaps scientific, philosophical or even meteorological reasons. However, the well-known 'Arabic' record of the SN 1054 sighting was indeed of European origin as the writer, ibn Butlan, was in Constantinople at the time when he observed it, but only later reported it when he was safely away from the reach of the treacherous Byzantine Empire.

Certainly, there are no other precise and indisputable European records of SN 1054 that are comparable to the ones from the East-Asian countries. While there is no doubt that most (if not all) of the historical records around SN 1054 suffer from various biases (from which temporal coincidence is the most dominant), we also explored a plausible explanation of some European records on tombstones from Bosnia and Herzegovina that could relate directly to the SN 1054 event.

Yet this does not explain the almost total absence of European archival documentation of the supernova. Perhaps the most plausible explanation for this would be the poor scientific knowledge (and interest) of celestial events such as supernovae at this time, and the added likelihood that the Christian Church ignored the event. Given the Church's stand on astronomy/astrology, there would be a strong incentive not to report the occurrence of any event - including an obvious supernova - that would threaten the theological/astronomical status quo.

Here, we explore another idea that SN 1054 was in fact recorded in Constantinople, this time by means of a cipher.

## 2. Was SN 1054 recorded in secret?

In AD 1054, the SN explosion would have been one of the most obvious and intriguing astronomical events witnessed in the sky. Today we know this event through its remnant, the Crab Nebula, one of the most spectacular objects studied by astronomers (see Figure 1, in [1]). If you wanted to observe and study this object and record it for posterity, but were afraid of the consequences that would follow, how could you hide your discovery (which in this case was obvious anyway) but report it secretly so that it could be read and accurately deciphered at a later date? Yet, it is also conceivable that coins were used as a means of mass communication to project power (see https://www.govmint.com/coin-authority/post/history-of-dates-on-coins for a quick discussion of this).

Perhaps one of the ways for a clever astronomer at Constantine IX's University of Constantinople to record the event would be to use a cipher, in this case, a minted coin (Figure 1) of a special edition that were minted after the 1054 event. We make the hypothesis that the head of the Emperor, Constantine IX, displayed in a special edition of coins might symbolise the Sun with a bright 'star' on either side - Venus in the east and SN 1054 in the west, perhaps also representing the newly split Christian churches.

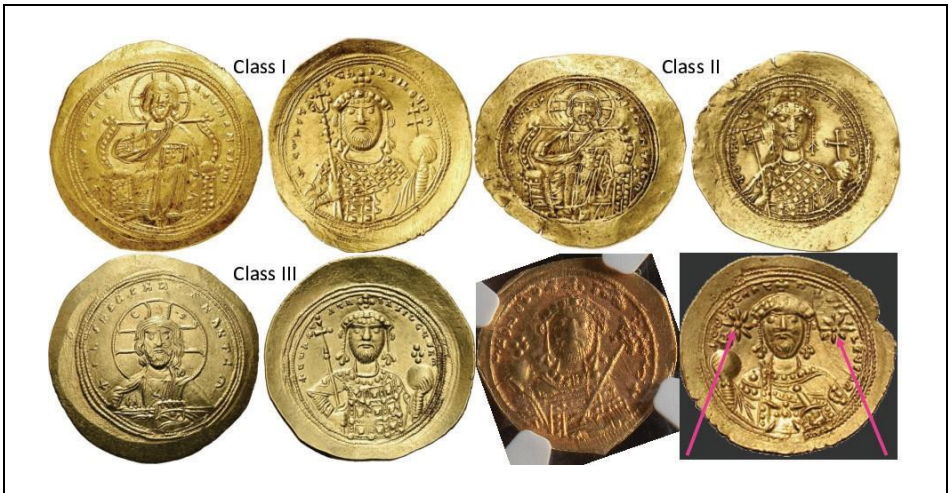

**Figure 1.** The four coinage classes of Constantine IX Monomachos minted from 1042 to 1055. The coin at the bottom right is Class IV. We postulate that for many days in northern summer of 1054 A.D., Venus and SN 1054 occupied diametrically opposite positions with respect to the Sun, as indicated by the arrows on the coin (where Constantine IX's head is suggested to represents the Sun). At the bottom, third from the left, is a rare and previously unclassified Constantine IX coin with a flat flan. Images credit: Classical Numismatics Group [CNG, www.cngcoins.com].

During the 4598 days of his rule, Constantine IX minted a few million coins in four known series [3-7]. Interestingly, in the fourth and last series (hereafter referred to as Class IV), Constantine IX's head appears between two stars [6], each with eight pointers (radii or rays). As pointed by Stephenson and Green, Chinese astronomers noted that SN 1054 "…had pointed rays on all sides" [8]. In this case, the stars are likely to reflect contemporary events of great significance, since they appear in only one specific (and most likely small number-limited) series. It is also quite possible that these coins were not for commercial use but were strictly commemorative.

In Judeo-Christian culture an eight-pointed star represents the beginning, resurrection (regeneration), salvation and super-abundance. The number eight is also associated with the arrival of Christ as a restorer of the process of creation [9]. Perhaps, this also pointed to the long-awaited Second Coming of Christ predicted in Revelations and expected for the year 1000, as suggested by Pope Sylvester II (946-1003) [10, 11]. This choice of eight rays also could be linked to the biblical account of the day of circumcision for the infant Jesus, and may have directly represented Constantinople, in the same way that the six-pointed star represented Thessaloniki [12]. If true, this would clearly suggest that the minting place of this special series was Constantinople. For numismatists, the eight-pointed star symbol on coins represents the incarnation of Christ and must have become publicly significant and almost exclusively used during specific random and rare astronomical events or alignments, such as supernovae, novae, impressive naked eye comets or planetary, stellar and lunar conjunctions [13]. Such symbolism of the eight-ray star could have been used to hide (or simply to be a discreet form of recording) the SN from the conservative Church at that time.

In Judeo-Christian culture an eight-pointed star represents the beginning, resurrection (regeneration), salvation and super-abundance. The number eight is also associated with the arrival of Christ as a restorer of the process of creation [9]. Perhaps, this also pointed to the long-awaited Second Coming of Christ predicted in Revelations and expected for the year 1000, as suggested by Pope Sylvester II (946-1003) [10, 11]. This choice of eight rays also could be linked to the biblical account of the day of circumcision for the infant Jesus, and may have directly represented Constantinople, in the same way that the six-pointed star represented Thessaloniki [12]. If true, this would clearly suggest that the minting place of this special series was Constantinople. For numismatists, the eight-pointed star symbol on coins represents the incarnation of Christ and must have become publicly significant and almost exclusively used during specific random and rare astronomical events or alignments, such as supernovae, novae, impressive naked eye comets or planetary, stellar and lunar conjunctions [13]. Such symbolism of the eight-ray star could have been used to hide (or simply to be a discreet form of recording) the SN from the conservative Church at that time.

If this was a hidden way of commemorating the appearance of SN 1054 then this final fourth series of coins must have been minted before Constantine IX's death on 11 January 1055 and over a very short period of about six months [7, p. 26]. Using the indictional cycles (any of the years in a 15-year cycle used to date medieval documents throughout Europe, both in the East and the West) [7, p. 27] dates the Constantine IX Class IV gold histamenon from 1 September 1053 to 11 January 1055. He also suggests that this could be a substantial coinage with ~500 original dies for obverse and reverse. We emphasise that all these estimates are based on the extant specimens (or subclasses). If one applies a conservative estimate of 10 000 coins per die, in 6.5 months some 1-2 million histamena of the Class IV alone could potentially have been minted (in the same years gold tetartera were also issued in the millions). We emphasise that this might be an upper limit for 'ordinary' class coins but very unlikely for this special and, as argued below, limited batch, which represent only ~5-6% of all known Constantine IX histamena.

We also question the claim by [4, 6] that only 20 of these coins are minted, for we were able to identify 36 such coins, and these are shown in Figure 2. Moreover, [7, p. 141] suggests that nearly 90 exist with a similar number of corresponding dies.

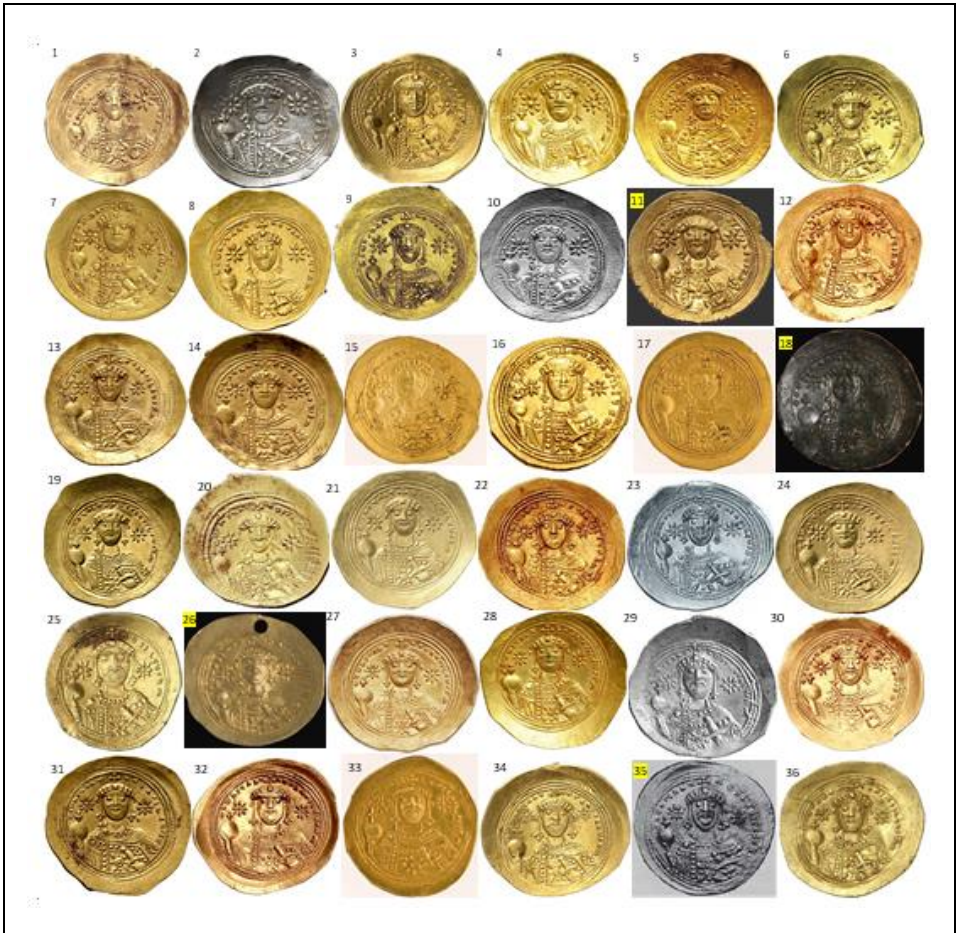

**Figure 2.** The coinage (reverse side) of Constantine IX Monomachos, minted in 1054-1055 (Class IV). The coins are sorted from the one where the stars are the largest (#1) to the smallest (#36). Images credit: Classical Numismatics Group (CNG, www.cngcoins.com).

### 2.1. A sample of thirty-two Class IV coins of Constantine IX

We investigated the 36 identified Constantine IX Class IV coins in various private collections and museums. We found obvious and very significant differences among each of them. These coins are quite small (typically ~25 mm in diameter) and they are cup-shaped (a.k.a. scyphates). [10] argues that perfectly struck scyphates are very uncommon. From Figure 2 we can clearly see those three different types of the Emperor's clothing are represented (for example coins #5, #6 and #7), implying that these coins were made with different dies and most likely at different times. Closer inspection of each coin reveals that there are at least six subclasses of this Class IV coin (Table 1, Column 2). These classifications are based on differences relating to the exact size and shape of the 36 coins that we investigate here.

**Table 1.** Properties of 36 known Emperor Constantine IX coins of Class IV. The coin numbers in column 1 correspond to those in Figures 2 and 3. We introduce six subclasses in column 2 based on shapes of the stars and their overall appearance of the coins. SL stands for 'star left' and SR stands for 'star right'. SD is standard deviation of the mean. The image resolution (column 9) describes the quality/resolution of each coin as high (HR), standard (SR) or low (LR). Column 10 is Venus apparent magnitude [https://ssd.jpl.nasa.gov/horizons.cgi] for determining proposed coin minting date (Column 11).

| Coin # | Subclass Type | $SL_{maj}$ (mm) | $SL_{min}$ (mm) | $SR_{maj}$ (mm) | $SR_{min}$ (mm) | Stars$_{avg}$ (mm) | Stars$_{avg}$ % | Image Resolution | Venus mV | Proposed Date(s) |
|---|---|---|---|---|---|---|---|---|---|---|
| (1) | (2) | (3) | (4) | (5) | (6) | (7) | (8) | (9) | (10) | (11) |
| 1 | 1b | 4.00 | 4.00 | 4.15 | 4.85 | 4.25 | 100.0 | HR | −4.278 | 16/07/1054 |
| 2 | 4a | 4.00 | 4.00 | 4.50 | 4.00 | 4.125 | 97.1 | HR | −4.365 | 25/07/1054 |
| 3 | 3 | 4.50 | 4.00 | 4.00 | 3.50 | 4.00 | 94.1 | HR | −4.180 | 04/07/1054 |
| 4 | 4a | 4.00 | 4.00 | 4.00 | 4.00 | 4.00 | 94.1 | HR | −4.180 | 04/07/1054 |
| 5 | 4a | 3.75 | 4.00 | 4.00 | 4.00 | 3.9375 | 92.6 | HR | −4.385 | 27/07/1054 |
| 6 | 3 | 3.75 | 4.00 | 3.75 | 4.00 | 3.875 | 91.2 | HR | −4.417 | 30/07/1054 |
| 7 | 2 | 3.50 | 4.00 | 3.50 | 4.00 | 3.75 | 88.2 | HR | −4.449 | 02/08/1054 |
| 8 | 1b | 3.75 | 3.75 | 3.75 | 3.50 | 3.6875 | 86.8 | HR | −4.482 | 05/08/1054 |
| 9 | 1b | 3.75 | 3.75 | 3.50 | 3.50 | 3.625 | 85.3 | HR | ∼ −4.5 | 09–13/08/1054 |
| 10 | 3 | 3.25 | 3.75 | 3.50 | 4.00 | 3.625 | 85.3 | SR | ∼ −4.5 | 09–13/08/1054 |
| 11 | 4a | 3.30 | 3.70 | 3.50 | 3.50 | 3.50 | 82.4 | SR | −4.6 to −4.8 | 15/08–09/1054 |
| 12 | 4a | 3.50 | 3.50 | 3.50 | 3.50 | 3.50 | 82.4 | HR | −4.6 to −4.8 | 15/08–09/1054 |
| 13 | 4b | 3.50 | 3.50 | 3.50 | 3.50 | 3.50 | 82.4 | LR | −4.6 to −4.8 | 15/08–09/1054 |
| 14 | 4a | 3.25 | 3.50 | 3.50 | 3.75 | 3.50 | 82.4 | HR | −4.6 to −4.8 | 15/08–09/1054 |
| 15 | 3 | 4.00 | 3.00 | 3.50 | 3.50 | 3.50 | 82.4 | SR | −4.6 to −4.8 | 15/08–09/1054 |
| 16 | 3 | 3.50 | 3.50 | 3.25 | 3.75 | 3.50 | 82.4 | HR | −4.6 to −4.8 | 15/08–09/1054 |
| 17 | 4a | 3.25 | 3.75 | 3.50 | 3.50 | 3.50 | 82.4 | SR | −4.6 to −4.8 | 15/08–09/1054 |
| 18 | 3 | 3.25 | 3.50 | 3.50 | 3.50 | 3.4375 | 80.9 | LR | −4.740 | 20/09/1054 |
| 19 | 4a | 3.75 | 3.50 | 3.25 | 3.25 | 3.4375 | 80.9 | HR | −4.740 | 20/09/1054 |
| 20 | 4a | 3.30 | 3.70 | 3.10 | 3.40 | 3.375 | 79.4 | HR | −4.369 | 30/09/1054 |
| 21 | 4b | 3.25 | 3.50 | 3.25 | 3.25 | 3.3125 | 77.9 | HR | −4.208 | 10/10/1054 |
| 22 | 4a | 3.00 | 3.50 | 3.25 | 3.25 | 3.25 | 76.5 | SR | −4.2 to −4.8 | 11–30/10/1054 |
| 23 | 3 | 3.25 | 3.25 | 3.25 | 3.25 | 3.25 | 76.5 | LR | −4.2 to −4.8 | 11–30/10/1054 |
| 24 | 4b | 3.25 | 3.50 | 3.125 | 3.125 | 3.25 | 76.5 | HR | −4.2 to −4.8 | 11–30/10/1054 |
| 25 | 3 | 3.00 | 3.25 | 3.50 | 3.00 | 3.1875 | 75.0 | HR | −4.900 | 10/11/1054 |
| 26 | 4b | 3.00 | 3.00 | 3.50 | 3.10 | 3.15 | 74.1 | HR | −4.862 | 20/11/1054 |
| 27 | 1a | 3.00 | 3.25 | 3.00 | 3.30 | 3.1375 | 73.5 | HR | −4.818 | 25/11/1054 |
| 28 | 4b | 3.00 | 3.00 | 3.10 | 3.40 | 3.125 | 73.5 | SR | −4.757 | 01/12/1054 |
| 29 | 3 | 3.00 | 3.00 | 3.25 | 3.50 | 3.0625 | 80.9 | SR | −4.7 to −4.6 | 5–15/12/1054 |
| 30 | 4a | 3.00 | 3.25 | 3.00 | 3.00 | 3.0625 | 72.1 | HR | −4.7 to −4.6 | 5–15/12/1054 |
| 31 | 4a | 3.00 | 3.25 | 3.00 | 3.00 | 3.0625 | 72.1 | HR | −4.7 to −4.6 | 5–15/12/1054 |
| 32 | 4b | 3.00 | 3.00 | 3.00 | 3.00 | 3.00 | 70.6 | HR | −4.546 | 20/12/1054 |
| 33 | 1b | 3.00 | 3.00 | 2.75 | 3.00 | 2.9375 | 69.1 | SR | −4.492 | 25/12/1054 |
| 34 | 4a | 3.00 | 3.00 | 2.75 | 2.75 | 2.875 | 67.6 | HR | −4.382 | 05/01/1055 |
| 35 | 3 | 2.75 | 2.75 | 2.75 | 2.75 | 2.75 | 64.7 | LR | ∼ −4.3 | 10–15/01/1055 |
| 36 | 4b | 2.75 | 3.00 | 2.50 | 2.75 | 2.75 | 64.7 | SR | ∼ −4.3 | 10–15/01/1055 |
| Mean | | 3.36 | 3.47 | 3.37 | 3.44 | 3.41 | 80.3 | | | |
| SD | | 0.41 | 0.37 | 0.43 | 0.44 | 0.38 | 8.9 | | | |

*2.1.1. Does the cipher contain an inadvertent or attempted light curve?*

As noted above, [10] suggested that the size of the stars on each coin might be different. We tested this hypothesis by measuring the sizes of the two stars on each of 36 available Class IV Constantine IX coins (Table 1). We noted that all coins showed wear from ~966 years ageing, and that most of them were deformed in various ways, which prevented us from taking even more precise measurements.

The Harvard Art Museums scanned two of these 36 coins (Figure 3) [https://www.harvardartmuseums.org/], at the same time and next to the coin they scanned a ruler that provided us with a calibration point for the rest of our sample (Figure 2).

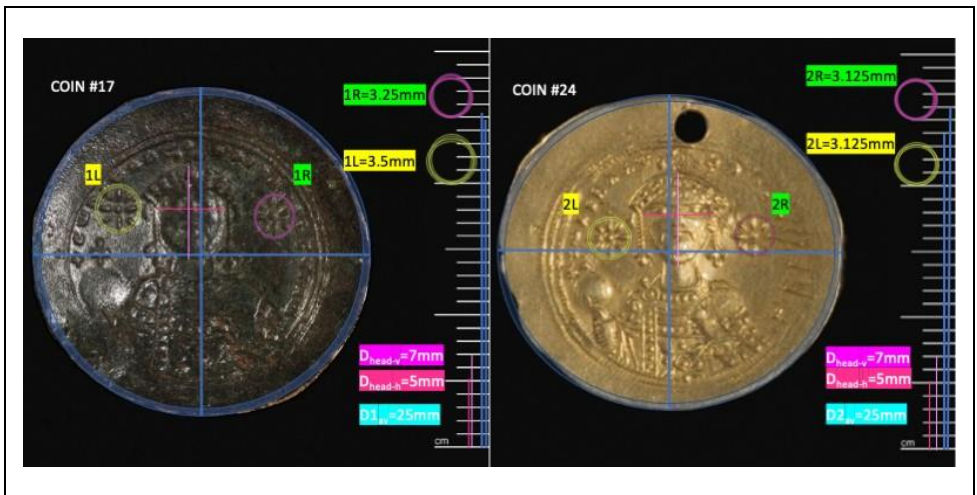

**Figure 3.** The two coins of Constantine IX Monomachos (Class IV) scanned by the Harvard Art Museums thatwe used as calibrator coins for our study [https://www.harvardartmuseums.org/].

We focused our calibration measurements on the central part of the two coins - the Emperor's head - which measured 7 and 5 mm in size (vertical and horizontal), as shown in Figure 3. The choice of using the Emperor's head was based on our understanding of the production process [Markowitz, 2014, https://coinweek.com/education/], where the central part of the coin is most likely to preserve its original and intended size and shape. The side ruler and scanned higher resolution images allowed us to set our measurement error at 0.25 mm, which was one quarter of the smallest ruler division of 1 mm. Also, we cross-checked the size of the Emperor's head in other Constantine IX coin classes and found no significant discrepancies. We then applied this to the remaining 34-coin images (shown in Figure 2), where minor adjustments (e.g. rescaling) were applied on only a handful of images. Finally, each star was fitted with an ellipse and its major and minor axes were measured and recorded in Table 1 (Columns 3, 4, 5 and 6). Column 7 in Table 1 lists the average star size for each coin, while Column 8 shows the percentile change from the largest star on coin #1.

While there are no exact historical records, we hypothesise that the coins were intentionally minted in order of decreasing size of the stars. Very surprisingly, we found that the stars in these 36 coins gradually decreased in size ('faded') by ~35%; from the largest (4.25 mm; 100%) to the smallest (2.75 mm or 64.7%; see Table 1). While there is obvious observable change in size, a standard deviation of 0.37 to 0.44 across this sample of 36 coins implies that it is hardly significant. In Figure 2 we show the whole sample of Constantine IX coins (Class IV only) sorted by the size of the stars from the largest (#1) to the smallest (#36). Note that the calibrator coins #18 and #26 are also shown in Figure 3.

Further assurance about the real change in star size was obtained by using the same methodology but applied to very similar coins from the reign of Alexius I. While the sample of Alexius I coins with the stars are small (only 4), we found no change in star size at all. So, the question is: can this large (1.50±0.25 mm) decrease in the size of the stars on Constantine IX Class IV coins be attributed to intentional ciphering, or is it the product of a non-perfect minting production process, or did we select a non-random sample for our study?

To examine these questions, we began by examining imperfections in XI century gold coin production. From the high-resolution images of the Constantine IX coins we could clearly see that the imprint of the Emperor's face had numerous submillimetre details - well below our above assumed measurement error of (0.25 mm). But, could the craftsmen control such precision during the production process, as described by Markowitz [https://coinweek.com/education/]? We suggest that such control and minting perfection would not have been a trivial task but quite possible. The large diversity of details shown on almost all of the 36 Class IV coins points towards a very special line of production that specifically relied on a one-coin-one-mold philosophy. We emphasise that we did not find such diversity among the other three significantly larger samples of Constantine IX coin classes. This suggests that the production of the Class IV coins would not have been a cheap, quick or easy exercise if the intention was to depict the SN 1054 event. Or perhaps the survival rate of these Class IV coins was such that they exceeded usual rules as suggested by [7, p. 26]?

If we assume that a population of as small as ~150-200 Constantine IX Monomachos Class IV coins was connected to SN 1054 then this could imply that their production was spread over a period of about six months (from July 1054 until January 1055), that is, about one Class IV coin per day. Further, assuming that the intentional production of a limited number of Class IV coins with an on-going gradual reduction in the size of the stars is correct, then we can ask why was this done? Did the conspirators want the stars on these coins to reflect the actual fading of SN 1054? Was the mould based on what the craftsman or a supervisor saw in the sky by intention or was it by accident? Or is this all simply the result of a random sample of different size coin stars? And is there any way to test this hypothesis?

As a morning day-time object, Venus was at $m_v = -4.278$ mag on 16 July 1054 (Table 1, column 10). The ephemeris [https://ssd.jpl.nasa.gov/horizons] accurately predicted that in the northern summer of 1054 the apparent magnitude of Venus was between -3.9 (in April 1054) and -4.9 (in November 1054). Also see [11]. As mentioned earlier, since Chinese astronomers could no longer see it in the daytime after 23 days, it is reasonable to assume that at its peak SN 1054 was apparent visual magnitude -4.3 (on 16 July 1054). However, we caution that the brightness curves of Venus and SN 1054 would be very different as can be seen in Figure 4. SN 1054 was a most likely core collapse type IIn-P event as suggested by [14] with a resultant neutron star pulsar (i.e. the Crab Nebulae and Pulsar). Based on well-established East-Asian records (Table 2) and compared to well-studied SN events of a similar type, one can expect that SN 1054 would have a sharp rise in early July of 1054 with reaching a peak brightness on 16 July 1054. This would be followed up by somewhat faster decline until mid-August 1054 (to $m_V = -3.5$) after which more shallow decay would be evident. These types of SN have been reported to have an average decay rate of 0.008-0.0075 magnitudes per day [15] and one can estimate a decline of $m_V \sim$ 1.4-1.5 magnitude over the period from mid-July 1054 to mid-January 1055 (190 days $\times$ 0.0075). This implies that the apparent brightness of SN 1054 could fell from a peak of $m_V \sim$ -4.3 mag in July 1054 to $m_V \sim$ -2.9-2.8 mag in January 1055. This would be strikingly similar to the changing size of the stars on our Class IV coins (coins decreased in size from 4.25 to 2.75 mm, Table 1, column 8). The apparent magnitude of type IIn-P SN 1054 also fell at a similar rate of $\sim$35%.

**Table 2.** Historical and suggested timescale of SN 1054 events.

| Date | SN 1054 $m_V$ | Comment |
|---|---|---|
| 4 July 1054 | -4.0 | East-Asian record; naked eye sightings during day-time |
| 16 July 1054 | -4.3 | East-Asian record, peak of brightness |
| 27 July 1054 | -4.1 | East-Asian record, the end of day-time recognition (23[rd] day after initial day-time detection) |
| 15 August 1054 | -3.5 | Here estimated, end of the sharp decline in brightness |
| 11 January 1055 | -2.8 | Death of Constantine IX |
| 6 April 1056 | +5.5 | East-Asian record, the last sighting |

In Figure 4 we show the second half of 1054 with rise and decline of SN 1054 brightness (green curve) as well as the apparent brightness of Venus (blue curve). If these were intentional exercises, one can assign possible dates of minting to each of the 36 coins (gold vertical lines and shades) as listed Table 1 (column 11). These are suggested based on an apparent size of stars on these coins for which we investigate possible associations. While, once again, we emphasis that no firm written records are available to support precise dating on any of these Constantine IX coins of Class IV, one can see striking resemblance with SN 1054 possible type IIp-n brightness curve.

Could the changing size of the stars on the coins represent a deliberate attempt to reflect the declining magnitude of SN 1054? This raises the question of whether the astronomers of the time could observe such small changes in magnitude. Schaefer (pers. comm. 2018) is not only sceptical about the accuracy of the East-Asian records, but he also questions the ability of astronomers/astrologers in 1054/1055 to observe this gradual change in stellar brightness, yet it is well known that these days experienced variable star observers can consistently distinguish magnitude variations of 0.1. Also, the knowledge and especially recognition of the daytime star is certainly not the 'everyday' job of the scientist at the time of SN 1054 event (for a review about the visibility of daylight stars see: [16-20]). Further on, there are also no written records that medieval astronomers observed a change in Venus's brightness. Without more information about the dates of minting of these coins, we will never know if there was any intention of creating a light curve.

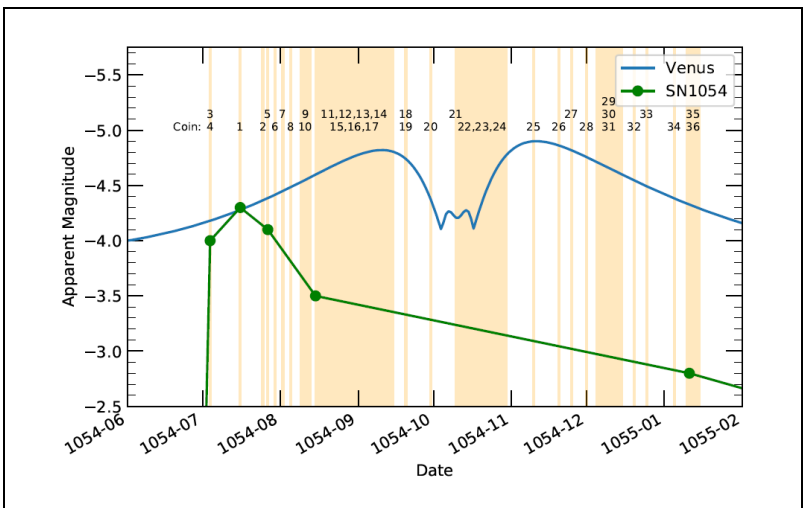

**Figure 4.** Apparent brightness of Venus (blue curve) and SN 1054 (green curve) in the second half of 1054. SN 1054 light curve estimate is based on assumption of SN type IIn-P event and reconstructed historical records shown in Table 2. Vertical lines represent possible minting dates of Constantine IV Class IV coins (Table 1, columns 1 and 11).

Among others, the most interesting is the unique coin #7 (Figure 2) where we see a mixture of various elements and even possible double minting or/and post-minting handcraft intervention. We note that the only part of these Class IV coins that does not change in any shape or form is the head of Emperor Constantine IX. This could be attributed to the newly developed technique of minting such concavo/convex shape coins as described by [21].

Another way to check the rarity (or limited edition) of this Class IV coin is to search for its present-day market (auction) value. The specialised coin search site ACSEARCH.INFO (also see CoinArchives.com) finds about 5000 sales/offers of Constantine IX Monomachos gold coins over the past 17 years. Of these 5000 only 15 (0.3%) were of Class IV. The top and most recent sale of Class IV coin was ~$10000 US while other Constantine IX Classes barely reach $100 US (about 100:1 ratio). Given that the Constantine IX minted millions of gold coins over his period of rule and that this particular Class IV would be time limited compared to other coin Classes to <10% fraction Constantine IV gold coins, one could speculate that this Class IV limited (or special) edition might have comprised from 150-200 to, at most, a few thousand coins - which is a marginally smaller number than [7, p. 26] finds in his very extensive search.

Conspicuously, [10] found an additional two of these Constantine IX Class IV coins to have their stars carefully removed (scratched off or altered) which further adds to the mystery. Among several possibilities considered by [22], we suggest that conservative Christians may not have been impressed with this 'cipher' and decided to remove it which may imply that this was a known secret. We also must wonder if our 36 coins examined here are representative of the assumed total population of Constantine IX Class IV coins. In other words, is our sample 15-20% of the population? This is especially interesting given that we do not have clear evidence for dating of any of these coins.

Stars were used in a number of occasions (for example: Justinian I the Great (r. 527-565), Anastasius I (r. 491-518), Maurice Tiberius (r. 582-602), Heraclius (r. 610-641) and Justinian II (r. 685-711)) on Eastern Roman Empire coins back in the VI and VII centuries AD, primarily as a symbol to indicate a value of 23 carats [23]. We also found that the few of these same rulers issued silver coins with the eight-ray star, which is in contradiction with the previous claim of a star being a symbol of 23 carat value. However, our particular Constantine IX coins of Class IV are known to be well debased [3, 7, 24, 25] valued them to be only 19.5 carats (the lowest of all four Constantine IX series). If correct, it is very unlikely that in the XI century there would have been any memory of this earlier practice. We also stress here that there is no written evidence that the Eastern Roman Emperors used iconography to indicate the degree of debasement of a particular coin series.

Constantine IX Class IV coins are not the only coins from the XI century AD with the same or similar appearance [26]. For example, [22] noted that similar symbols (stellati or 'with stars') were found on another 30-31 Eastern Roman Empire coins, and that they usually had religious meanings. We note that two stars appear next to the Emperor's figure on 18 of these 30-31 coins, but more importantly, not all of these are positioned next to the Emperor's head. The coins of the next two Emperors, Michael VII Ducas (r. 1071-1078) and Alexius I (r. 1081-1118) also had two stars depicted, one on either side of the Emperor's head [http://www.wildwinds.com/, Triton V, 2310], but no obvious astronomical event can be linked to this period. Similarly, [7, p. 24] points to stars in the electrum trachy from John III Ducas (1246-1254). No obvious astronomical event can be linked with this period either. A notable difference of Michael VII and John III coins is that two stars are minted on either side of Christ (obverse side) not the emperor (reverse side). However, a single 'star' on an electrum aspron trachy (silver or silver alloy coins) of Alexius I [http:/www.coin archives.com/; D.O.IV23c] may refer to a comet from 1105/1106 [27, 28]. Similarly, famous William I (the Conqueror; 1028-1087) coins from 1067 and 1076 also show two stars that represent two bright comets, one of which was 1P/Halley [26, p. 63].

Because of the above, Schaefer (pers. comm., 2018/2019) argues that the Constantine IX stellati are just an ordinary part of a long tradition throughout all East Roman Empire times to place a pair of stars on either side of the Emperor's figure on the obverse of their coinage. As such, Schaefer insists that the star pair on the stellati might have nothing to do with any supernova [22].

Kramer suggests that the star, while still in effulgence, represents a clergy (patriarch) who has not lost his senses through heresy [29]. Specifically, he suggested that the SN 1054 would symbolise a lapse into schism and retention of all true doctrines. The star fading ('burning') is explained as an allusion to the teachings of the illustrious intellectuals in Constantinople.

Dimitrijevic has interpreted stars on medieval coins as primarily a source of light [30, 31]. Furthermore, coins could have a heavenly meaning to the church by depicting the conflict between the spiritual forces or light and the material forces or darkness. However, one must be cautious about applying such an opinion to medieval Europe as a whole, where stars on coins could have been symbols for any number of things, or perhaps for nothing at all.

Minting a star (or stars) on many Roman and Greek coins was a well-adopted practice and usually had a reference to astrology [32-34]. For example, the first Roman Caesar Augustus coin commemorated the Great Comet of 44 BC. Among the pagan Romans, stars were seen as a symbol of eternity or dedication to a special purpose or service (usually religious). At the same time, they were a sign of glory and were frequently used as mint-marks. Might these successive rulers also have adopted the cipher of SN 1054?

## 3. A cover up or not?

Considering everything we have discussed thus far, we come full circle to our initial question, 'Why are thereso few (if any?) clear documentary records of SN 1054 from Europe, especially since 48 years earlier a similar event (SN 1006) is well documented?' We now briefly summarise some possible reasons.

If scientists present at the time were unwilling to investigate this event further, perhaps it was because of a philosophical prejudice against any observed changes in the supposedly perfect and eternal night sky [35]. The chaotic events surrounding the serendipitous SN 1054 may have caused the hierarchy to decide it would be prudent to simply ignore this particular celestial event. But, what about astronomers elsewhere in Europe (and England)? Why did they fall over?

If observations of SN 1054 were recorded, perhaps they did not withstood the destructive effects of time. There were no records from 'specialised' Eastern Roman Empire sky-watchers (excluding perhaps a few Monks from the central Europe) like Chinese Imperial Astrologers who kept daily diaries, thus limiting the number of possible surviving documents. Records from the Eastern Roman Empire had to survive several brutal medieval wars and various changes of governments (and religion). The people of the time may not have had plans for the protection or evacuation of such records and during these 'Dark Ages', and the largest number of surviving written documents are primarily liturgical/religious writings. In fact, Holtzmann references 1089 negotiations with the papacy, but he could not find any corresponding documents from the Patriarch Cerularius [36]. What happened to these records?

Psellos and ibn Butlan, were in Constantinople immediately before, during and after the SN 1054 event and had close financial connections with the Emperor and the Patriarch. Once ibn Butlan left his well-paid post in Constantinople, he announced SN 1054 from the safety of Cairo. Interestingly, in the autumn of 1054 Psellos became a monk (a usual ploy to avoid jail or death sentences [37]), and he moved to Mount Olympus [38]. He wrote that this decision was dictated by political affairs (the Emperor's inconsistency) and later left the impression that his reasons were partly astrological [39]; Psellos clearly stated: "…I am no believer in the theory that our human affairs are influenced by the movements of the stars ... [and] I certainly do not believe that the positions or the appearance of stars affect what goes on in the sublunary world." [39, p. 35, 76]

Was Psellos suggesting that his views on astrology/astronomy collided with the official state/church doctrine at the time, and that in the fall of 1054 astrology was perhaps banned or discouraged? It is interesting that in 1056, Psellos abandoned monastic life and returned to serve the court of the Eastern Roman Empire. We note that the SN 1006 event was recorded when Persian astronomers Ali ibn Ridwan (ca AD 988-1061) and Ibn Sina [Lat. Avicenna, AD 980-1037; 40] - who also had studied in Constantinople - announced the SN.

Surely, ibn Butlan and Psellos were not the only competent astronomers in all of Christendom in AD 1054, and one would expect that the numerous monastic societies found throughout Europe and the British Isles would have recorded this event. At least some of them would be outside the reach and control of Rome and Constantinople. Yet there were monks in central Italy and in Switzerland who recorded SN 1006 event [41, 42]. Additionally, a stone mason from Radimlja may have recorded SN 1054 [2].

[43-45] have suggested that perhaps the whole of the Middle East and Europe was clouded out for six months when SN 1054 occurred, but we argue that the crucial 3-4 weeks of bad weather in July 1054 cannot explain the lack of records. After all, it was the northern hemisphere summer and the weather would not have been continuously bad for more than a few days.

Similarly, Ridgway has suggested that an Icelandic volcano could have blacked out the sky during this period [46]. However, our search for any volcanoes in or close to Europe did not reveal any major or conclusive records of volcanic activity, apart from the poorly documented ~AD 1060 volcano in Vatnajökull in Iceland [47]. Meanwhile, the eruption of the Baitoushan volcano in China in AD 1054 was one of the largest eruptions in the world in the past 10000 years [48], but this did not prevent Chinese, Korean and Japanese astronomers from recording the SN. However, several studies suggest that the actual date of this so-called 'Millennium Eruption' was actually AD 946 (e.g. see [49]).

## 4. Concluding remarks

We have surveyed the records of European history and culture from around 1054 to ask whether what we intuit is consistent with SN 1054 having been seen in the skies above Europe and having had an impact on the peoples who saw it. We have reviewed and analysed several factors that could account for the lack of clear European records of SN 1054. Possible explanations for this mystery include a political/religious conspiracy (a 'cover-up') or scientific, philosophical and/or meteorological reasons but its cryptic character prevents any conclusive claims.

## Acknowledgment

We thank Milan S. Dimitrijevic, Danica Draskovic, Dragoslav Komatina, Ray Norris, Wayne Orchiston, Bradley Schaefer, Velibor Velovic and the Western Sydney University Library team led by Linda Thornley for valuable help in reading and discussing various ideas regarding this study. Special thanks go to the National Museum in Belgrade for their help in obtaining photographs of coins.